\documentclass[aps,12pt,final,notitlepage,oneside,onecolumn,
nobibnotes,nofootinbib,amsmath,amssymb,
noshowpacs,centertags]%
{revtex4}

\newcommand{\bc}{\begin{center}}
\newcommand{\ec}{\end{center}}
\newcommand{\be}{\begin{equation}}
\newcommand{\ee}{\end{equation}}
\newcommand{\ba}{\begin{array}}
\newcommand{\ea}{\end{array}}
\newcommand{\bea}{\begin{eqnarray}}
\newcommand{\eea}{\end{eqnarray}}
\newcommand{\bt}{\begin{tabular}}
\newcommand{\et}{\end{tabular}}

\newcommand{\bsl}{\boldsymbol}

\begin{document}

\title{Charm and bottom baryons in nonperturbative quark
dynamics~\footnote{Contributed talk at the session of Russian
Academy of Sciences, {\it ``Physics of Fundamental
Interactions''}, ITEP, Moscow, November 26-30, 2007}}
\author{\firstname{I.~M.}~\surname{Narodetskii}}
\email{naro@itep.ru}

\author{\firstname{M.~A.}~\surname{Trusov}}
\email{trusov@itep.ru}

\author{\firstname{A.~I.}~\surname{Veselov}}
\email{veselov@itep.ru} \affiliation{Institute for Theoretical and
Experimental Physics}

\vspace{1cm}

%
%

\vspace{1cm}

\begin{abstract}
We use the field correlator method in QCD to calculate the masses
of $\Sigma_c$, $\Xi_c$ and recently observed $\Sigma_b$, $\Xi_b$
baryons and their orbital excitations.
\end{abstract}

\maketitle

\section{Introduction}
A comprehensive knowledge about the mass spectrum and spin
splittings of heavy baryons is important for our understanding of
quantum chromodynamics. The spectroscopy of $c$ and $b$ baryons
has undergone a great renaissance in recent years. New results
have been appearing in abundance as a result of improved
experimental techniques including information on states made of
both light {($u,\,d,\,s$)} and heavy $(c,\,b)$ quarks \cite{M07}.
Before 2007, the only one baryon with a $b$ quark, the
isospin-zero $\Lambda_b^0$, was known.  Now,
we have the isospin one $\Sigma_b$, $\Sigma_b^{*}$ baryons and
$\Xi_b$. The CDF Collaboration has seen the states
$\Sigma_b^{\pm}$ and $\Sigma^{*\pm}$ \cite{cdf_1},
while D\O~ \cite{do} and CDF \cite{cdf_2} have observed the
$\Xi_b^-$. The masses of these states are summarized in Table
\ref{tab:new_experiments}.

On theoretical side there are many results on heavy baryon masses
from different approaches including a number of quark model
variations \cite{Capstick:1986, Roncaglia:1995, Ebert:2005}, HQET
\cite{Jenkins:1996}, sum rules \cite{sum-rule} and lattice
calculations \cite{Mathur:2002}. Recently there have been several
theoretical papers on the masses of $\Sigma_{b}$, $\Sigma_{b}^{*}$
and $\Xi_b$ based on the modelling the color hyperfine interaction
\cite{lipkin,KKLR,Rosner-Lipkin}. In the present paper we use the
field correlator method (FCM) \cite{DS} to calculate the masses of
the $S$ wave baryons containing $c$ and $b$ quarks and orbitally
excited states that will be experimentally accessible in the
future. The same exercise has been applied previously to the
ordinary and cascade hyperons \cite{DNV}. Using FCM, substantial
work has been done in the heavy-light meson sector \cite{KNS}.
However, so far only very few results have been reported for
baryons. A further study of charmed and bottom baryons using FCM
therefore seems worthwhile.

The dynamics of the $ud$ pair plays a relevant role, being mainly
responsible for the spin splitting in the strange sector. A
similar contribution is expected for charmed and bottom baryons.
Estimates of the one-pion exchange contribution to the baryon mass
give -180 MeV both for $\Lambda$ and $\Lambda_b$. Because our
approach misses the chiral physics effects we calculate in this
work the masses of the $\Sigma$ and $\Xi$ states which are
affected by the chiral dynamics only slightly. Note that the
$b$-baryons are structurally identical to the $c$-baryons: only a
charmed quark is replaced by a beauty quark. Consequently, the
analysis of $b$ states and the results are only a variation of
what is found for the charmed systems.

\section{The formalism}

FCM provides a promising formulation of the nonperturbative QCD
that gives additional support of the quark model assumptions. The
key ingredient of the FCM is the use of the auxiliary fields (AF)
initially introduced in order to get rid of the square roots
appearing in the relativistic Hamiltonian. Using the AF formalism
allows to write a simple local form of the Effective Hamiltonian
(EH) for the three quark system
\begin{equation}
\label{eq:H} H=\sum\limits_{i=1}^3\left(\frac {m_{i}^2}{2\mu_i}+
\frac{\mu_i}{2}\right)+H_0+V,
\end{equation}
where $H_0$ is the kinetic energy operator, $V$ is the sum of the
string potential and a  one gluon exchange potential, $m_i$ are
the bare quark masses,
and $\mu_i$ are the {\it constant} AF which are eventually treated
as variational parameters. Such an approach allows one a very
transparent interpretation of AF: starting from bare quark masses
$m_i$ one arrives at the dynamical masses $\mu_i$ which appear due
to the interaction and can be treated as constituent masses of
quarks. The string potential considered in this work is \be
V_{Y}({\bf r}_1,\,{\bf r}_2,\,{\bf r}_3)\,=\,\sigma\,r_{min},\ee
where $\sigma$ is the string tension and  $r_{min}$ is the minimal
length corresponding to the Y--shaped string configuration

The mass $M_B$ of a baryon is given
by
\begin{equation}
\label{M_B} M\,=\,M_0\,+\,\Delta E_{HF},\end{equation}
where $\Delta E_{HF}$ is the spin correction,
\begin{equation}
\label{eq:M_B0}M_0\,=\,\sum\limits_{i=1}^3\left(\frac
{m_{i}^2}{2\mu_i\,}+
\,\frac{\mu_i}{2}\right)\,+\,E_0(\mu_i)\,+\,C,\end{equation}
$E_0(\mu_i)$ being the energy eigenvalue of the Shr\"{o}dinger
operator $H_0\,+\,V$, and $\mu_i$ are defined from the minimum
condition \be\label{eq:mc}
\frac{\partial\,M_0(m_i,\mu_i)}{\partial\,\mu_i}\,
=\,0.\end{equation} For the light quarks
($m_i\,\ll\,\sqrt{\sigma}$) $\mu_i\,\sim\,\sqrt{\sigma}(1+{\cal
O}(\alpha_s))$ while for the heavy quarks
($m_i\,\gg\,\sqrt{\sigma}$) $\mu_i\,\approx\,m_i$. In Eq.
(\ref{eq:M_B0}) $C$ is the quark self-energy correction which is
created by the color magnetic moment of a quark propagating
through the vacuum background field \cite{S2001}. This correction
adds an overall negative constant to the hadron masses:
\begin{equation} \label{self_energy}
C\,=\,-\frac{2\sigma}{\pi}\,\sum\limits_i\frac{\eta(t_i)}{\mu_i},\,\,\,\,\,t_i\,=\,m_i/\lambda_g,\end{equation}
where $\eta(t)$ is the known function \cite{S2001} and
$1/\lambda_g$ is the gluonic correlation length. We use
$\lambda_g\,=\,$ 1 GeV.

Taking the approach implemented in \cite{DNV}, the
spin-independent masses can be obtained from  (\ref{eq:M_B0}). We
solve the non-relativistic Schr\"odinger equation with the
confining and Coulomb interactions by the hyperspherical method to
determine the constituent quark masses $\mu_i$ and the zero-order
baryon masses $M_{0}$. Then we estimate HF splittings from the
perturbative color-magnetic interaction with account of the wave
function corrections.

\be\Delta\,E_{HF}\,=\,\sum\limits _{i<j}\,\frac{\bsl\sigma_i\,
             \bsl\sigma_j}{\mu_i\mu_j}\,\left(\frac{4\pi\alpha_s}{9}\,\langle\,
             \delta({\bsl r}_{ij})\rangle\,+\,\frac{\sigma\lambda_g^2}{4\pi}\,\langle r_{ij}\cdot
             K_1(\lambda_g
              r_{ij})\rangle\right)\label{eq:HF}
\ee The first term in (\ref{eq:HF}) is the standard color-magnetic
interaction in QCD \cite{deRuhula}, while the second term,
proportional to the string tension $\sigma$, was first derived in
Ref. \cite{S2002}.

We use the basis in which a heavy quark is singled out as quark
$3$ but in which the light quarks are still antisymmetrized. The
calculation of the spin matrix elements in (\ref{eq:HF}) is
straightforward for $J\,=\,3/2$, as the expectation value of each
${\bsl\sigma}_i\, {\bsl\sigma}_j$ is 1. For $J\,=\,1/2$~~
${\bsl\sigma}_1\, {\bsl\sigma}_2\,=\,1$, ${\bsl\sigma}_3\,
{\bsl\sigma}_1\,=\,{\bsl\sigma}_2\, {\bsl\sigma}_3\,=\,-2$ for
$\Sigma_q$ and $\Xi'_q$ while for $\Xi_q$~~ ${\bsl\sigma}_1\,
{\bsl\sigma}_2\,=\,-3,\,\, {\bsl\sigma}_3\,
{\bsl\sigma}_1\,=\,{\bsl\sigma}_2\, {\bsl\sigma}_3\,=\,0$.

The contact  interaction in (\ref{eq:HF})  requires the
calculation of the $\delta$ function expectation values. These
contact probabilities were calculated using 3-body wave functions
obtained by a hyperspherical method. E.g. for $L\,=\,0$ the square
of the baryon wave function at zero relative two quarks separation
is \be\langle\,\delta(\bsl{r}_{ij})\,\rangle_{L\,=\,0}\,=\,
\mu_{ij}^{3/2}\,\frac{4}{\pi^2}\,\gamma_0,\ee where \be
\gamma_0\,=\,\int\limits_0^{\infty}\, \frac{{
u}\,_0^2(x)}{x^3}\,dx\label{Delta}\ee is universal for all quark
pairs, $u_0(x)$ is the hyperradial function normalized as \be
\int\limits_0^{\infty}\, { u}\,_0^2(x)\,dx\,=\,1,\ee with \be
x^2\,=\,\sum_i\,\mu_i\,({\bsl r}_i\,-\,{\bsl
R}_{cm})^2\,=\,\frac{\mu_1\,\mu_2}{M}\,r_{12}^2\,+\,\frac{\mu_2\,\mu_3}{M}\,r_{23}^2\,+\,
\frac{\mu_3\,\mu_1}{M}\,r_{31}^2,\ee and \[
\mu_{ij}=\frac{\mu_i\,\mu_j}{\mu_i\,+\,\mu_j},\qquad
M=\mu_1+\mu_2+\mu_3.
\] Note that wave function corrections which influence the
hyperfine splitting between the different baryons tend to affect
$\gamma_0$ by only a few per cent: $\gamma_0\,=\,0.1207$ for
$nnc$, $0.1197$ for $nsc$, $0.1161$ for $nnb$, and $0.1153$ for
$nsb$ (in units GeV$^{\,3/2}$).

The second term in (\ref{eq:HF}) is expressed in terms of the
integrals \be \langle r_{ij}\cdot K_1(\lambda
              r_{ij})\rangle\,=\,\frac{16}{\pi\lambda_g}\,\int\limits_0^{\infty}u^2(x)\left(\int
              \limits_0^{\frac{\pi}{2}}\xi
              K_1(\xi)\sin^2\theta\cos^2\theta\,d\theta\right)\,dx,\,\,\,\,\,\,\,\,\xi\,=\,\frac{\lambda_g\,
              x\,\sin\theta}{\sqrt{\mu_{ij}}}~.\ee


\section{The results}
We employ some typical values of the string tension $\sigma$ and
the strong coupling constant $\alpha_s$ that have been used for
the description of the ground state baryons: $\sigma\,=\,$ 0.15
GeV$^2$ and $\alpha_s\,=\,$ 0.39. We neglect the mass difference
between $u$ and $d$ quarks, writing $n$ to stand for either $u$ or
$d$. We use the current light quark masses $m_n\,=\,7\,$ MeV and
the (slightly updated) strange quark mass $m_s\,=\,185$ MeV found
previously from the fit to $D_s$ spectra \cite{KNS}. However, our
predictions need an additional input for the bare quark masses
$m_c$ and $m_b$. These were fixed from the masses of $\Sigma_c$
and $\Sigma_b$, respectively, $m_c\,=\,$ 1359 MeV and $m_b\,=\,$
4712 MeV.

The result of the calculation of the $S$ wave states is given in
Table \ref{L=0}. In this Table we also present the dynamical quark
masses $\mu_n$, $\mu_s$ and $\mu_Q$ for various baryons. The
latters are computed solely in terms of the bare quark masses,
$\sigma$ and $\alpha_s$ and marginally depend on a baryon. We also
display the results obtained without the HF corrections. The
baryon masses are for the isospin averaged states. The result show
good agreement between data and theoretical predictions.

In the $\Xi_Q$ (with $Q$ standing for either $c$ or $b$) the light
quarks are approximately in a state with $S\,=\,0$, while another
heavier state $\Xi'_Q$ is expected in which the light quarks
mainly have $S\,=\,1$. Both have total $J\,=\,1/2$. The effect of
$\Xi\,-\,\Xi'$ mixing due  to the spin-spin interaction is
negligible \cite{KKLR}. There is also a state $\Xi_Q^*$ expected
with total $J\,=\,3/2$. The hyperfine splitting between $\Xi_c^*$
and $\Xi_c'$ is found to be 69 MeV that agrees with the
experimental value ($\sim$ 70 MeV) \cite{PDG}, while the predicted
mass difference $\Xi_b^*\,-\,\Xi_b'\,=\,$ 26 MeV agrees with the
finding of Ref. \cite{KKLR}. However, our perturbative
calculations do not reproduce the observed $\Xi_c'$ - $\Xi_c$ mass
difference. The large hyperfine splitting between axial and scalar
$ns$ diquarks is usually described by the smeared
$\delta$-function that  requires additional model-dependent
assumptions about the structure of interquark forces.

A similar calculations were performed for the P-wave
orbitally-excited states, see Table \ref{tab:L=1}.  Our basis
states diagonalize the confinement problem with eigenfunctions
that correspond to separate excitations of the light and heavy
quarks ($\bsl{\rho}$\,- and $\bsl{\lambda}$\,- excitations,
respectively). Excitation of the $\bsl{\lambda}$ variable unlike
excitation in $\bsl{\rho}$ involves the excitation of the ``odd''
heavy quark. For states with one unit of orbital angular momentum
between $Q$ quark and the two light quarks we obtain
$M(\Sigma_{c}) = 2832$ MeV, $M(\Xi_{c}) = 2867$ MeV,
$M(\Sigma_{b}) = 6132$ MeV, and $M(\Xi_{b}) = 6164$ MeV, while the
states with one units of orbital momentum between the two light
quarks are typically $\sim$ 100 MeV heavier. Note that zero order
results of Table \ref{tab:L=1} do not include the spin corrections
(which are smaller than those for the $S$-wave states) and the
(negative)  string corrections contributing into the masses of the
orbitally excited baryons \cite{sc}. Our preliminary analysis of
the latters shows that the string corrections  tend to decrease
the masses of the P-wave states by $\sim$  30 MeV. A more complete
analysis will be given elsewhere.

\section{Conclusions}
We have calculated the masses of heavy baryons systematically
using the FCM and the perturbative color-magnetic interaction.
There are two main points in which we differ from other approaches
to the same problem based on various relativistic Hamiltonians and
equations with local potentials. The first point is that we do not
introduce the constituent mass by hand. On the contrary, starting
from the bare quark mass we arrive to the dynamical quark mass
that appears due to the interaction. The second point is that for
the first time we calculate the hyperfine splitting with account
of the nonperturbative spin-spin forces between quarks in a
baryon. We find our numerical results to be in agreement with
experimental data and calculation in other approaches.\\[1cm]

This work was supported by the RFBR grant  06-02-17120.


\newpage
\begin{table}
\caption{The masses of bottom baryons observed by CDF and D\O~
collaborations.} \vspace{5mm}

\begin{center}
\begin{tabular}{c||cc}
\hline& Mass (MeV)& Collaboration
\\\hline\hline
$\Sigma_{b}^+$&$5808^{+2.0}_{-2.3}({\rm
stat.})\pm1.7({\rm syst.})$& \\
$\Sigma_b^-$&$5816^{+1.0}_{-1.0}({\rm stat.})\pm1.7({\rm syst.})$&\\
$\Sigma_{b}^{*+}$&$5829^{+1.6}_{-1.8}({\rm stat.})\pm1.7({\rm syst.})$&\raisebox{2ex}{CDF\cite{cdf_1}}\\
$\Sigma_{b}^{*-}$&$5837^{+2.1}_{-1.9}({\rm stat.})\pm1.7({\rm
syst.})$&\\\hline &$5774\pm11({\rm stat.})\pm15({\rm syst.})$&
D\O~ \cite{do}\\ \raisebox{1ex}{ $\Xi_b^-$}&~$5793\pm 2.5({\rm
stat.})\pm 1.7({\rm syst.})$& CDF\cite{cdf_2}\\\hline
\end{tabular}
  \label{tab:new_experiments}
\end{center}
\vspace{1cm}

\end{table}

\begin{table}[t]
 \caption{Heavy Baryons with $L\,=\,0$. The values of the bare
 quark masses used in this calculation are~~
$m_n\,=\,7$,~$m_s\,=\,185$,~$m_c\,=\,1359$,~and~$m_b\,=\,4712$
MeV. The underlined masses have been used to fix $m_c$ and $m_b$.
The dynamical quark masses $\mu_i$ are  defined by Eq.
(\ref{eq:mc}). } \vspace{5mm}

\centering
\begin{tabular}{ccccccccc} \hline\hline
Baryon&~~$\mu_n$& ~~$\mu_s$&~~$\mu_h$&$M_0$&~$\Delta\,E_{HF}^{\rm
(p)}$~&~$\Delta\,E_{HF}^{\rm (np)}$~&$M$
\\ \hline\hline
$\Sigma_c$~~~&~~470~~&&~~1455~~&~~2479~~&-19&-6&\underline {2454}\\
$\Sigma_c^*$~~&~~470~~&&~~1455~~&~~2479~~&30&13&2522\\
$\Xi_c$~~~&~~476&~~522&~~1458 &2519&-39&-20&2460\\ \hline
$\Sigma_b$~~~&~~509&&~~4749&5806&0&2&\underline{5808}\\
$\Sigma_b^*$~~&~~509&&~~4749&5806&+19&8&5833\\
$\Xi_b$~~~&~~514&~~615&~~4751&5844&-36&-17&5791\\
\hline\hline
\end{tabular}
 \vspace{1mm}

\label{L=0}
\end{table}
\vspace{1cm}


\begin{center}
\begin{table}[htb]
\caption{Masses of the heavy baryons from the present work and
other approaches and the comparison with experimental data (in
MeV).} \vspace{5mm}

\begin{tabular}{c|ccccccccc}
\hline\hline  & \cite{Capstick:1986} &\cite{Roncaglia:1995}&\cite{Jenkins:1996}&\cite{Mathur:2002}
& \cite{Ebert:2005}& \cite{sum-rule} &this work&exp\\

\hline \\  $\Sigma_c~~$&2440~~&2453~~&
&2452~~&2439~~&$2411^{+93}_{-81}$~~&2454~~&$2454\,\pm\,0.18$~~\\
$\Sigma^*_c$&~~2495~~&2520~~&
&2538~~&2518~~&$2534^{+96}_{-81}$~~&2522~~&$2518.4\,=\,0.6$~~\\
$\Xi_c$~~&& 2468~~& &2473~~&2481~~&$2432^{+79}_{-68}$~~&2460~~&$2467.9\,\pm\,0.4$~~ \\
\\
\hline\\

$\Sigma_b$~~&5795&5820&5824.2&5847&5805&$5809^{+82}_{-76}$&{
5808}&{
5808}\\
$\Sigma^*_b~~$&5805&5850&5840.0&5871&5834&$5835^{+82}_{-77}$&{
5833}&{ 5829}\\  $\Xi_b~~$&
&5810&5805.7&5788&5812&$5780^{+73}_{-68}$&5791&
$5774\,\pm\,20$\\
&&&&&&&&{ $5793\,\pm\,3$}\\
\hline
\end{tabular}
\end{table}
\end{center}
\begin{table}[t]
 \caption{Heavy Baryons. $L\,=\,1$. The $\bsl{\lambda}$ excitations  involve the excitation of the
``odd'' quark ($c$ for $\Lambda_c,\,\Xi_c$ or $b$ for
$\Sigma_b,\,\Xi_b$), while the $\bsl{\rho}$ excitations involve
the excitation of the $ud$ diquarks. The bare quark masses are the
same as in Table \ref{L=0}. } \label{tab:L=1} \vspace{1cm}

\centering
\begin{tabular}{ccccccc} \hline\hline\\
Baryon&~~${\bf L}_{\alpha}$ &~~$\mu_n$&~~ $\mu_s$&~~$\mu_h$&~~
$E_{0}$&$M$\\ \\ \hline\hline\\ $nnc$&${\bf
1}_{\rho}$&536&&1452&1397&2920\\  $nnc$&${\bf
1}_{\lambda}$&495&&1491&1377&2832
\\
$nsc$& ${\bf 1}_{\rho}$&542&582&1455&1372&2954\\
$nsc$& ${\bf 1}_{\lambda}$&497&544&1494&1353&2867\\
\\

\hline\\
$nnb$&${\bf 1}_{\rho}$&570&&4746&1294&6240\\  $nnb$&${\bf
1}_{\lambda}$&540&&4764&1234&6132
\\
$nsb$&${\bf 1}_{\rho}$&574&615&4748&1271&6272
\\
$nsb$&${\bf 1}_{\lambda}$&542&588&4765&1211&6164
\\ \\
\hline\hline

\end{tabular}
\vspace{1cm}

\end{table}
\end{document}